# An Overview of Numerical Simulations in Accretion Physics

Biplob Sarkar*; Liza Devi; Asish Jyoti Boruah
* Corresponding author
Department of Applied Sciences, Tezpur University, Napaam, Assam 784028, India
E-mail of the corresponding author: biplobs@tezu.ac.in

*Abstract*—Accretion physics studies the process of gravitational capture of ambient matter by massive stars. The background processes are very challenging to observe and measure due to the extreme conditions in these systems. Numerical simulations play a crucial role in accretion physics because they provide the only practical method to model the complex processes occurring in accretion disks. In this review, we outline different branches of numerical simulations, such as hydrodynamic simulations, magnetohydrodynamic simulations, and Monte-Carlo simulations, and their methodology, and we discuss possible implications for modeling accretion physics around black holes, neutron stars, and protoplanetary disks.

## I. Introduction

According to Horvath [1], first, an orbital dynamics problem motivated the research of the accretion process. The notion was initially established by the French astronomer E. Roche in 1873 and was associated with the shape of the inner equipotentials in a binary system [1, 2]. Although the concept is more generic and may be extended to a particle distribution in a gas, for example, Roche was concerned with the destiny of point satellites in orbit [1].

Mass accretion, the process by which a celestial body grows in mass by absorbing surrounding matter, is a crucial astrophysical process that releases gravitational potential energy, potentially making accreting objects extremely energetic and powerful sources of energy [3]. The accreting objects in the universe are many and varied, including protoplanets, protostars, black holes (BHs), and X-ray binaries (XRBs). Accretion often occurs as a disk. By studying accretion, we can extract information about the central compact accretor's characteristics (e.g., spin and mass). Accretion is an efficient way to extract gravitational energy, and it is the principal mechanism for generating high-energy radiation.

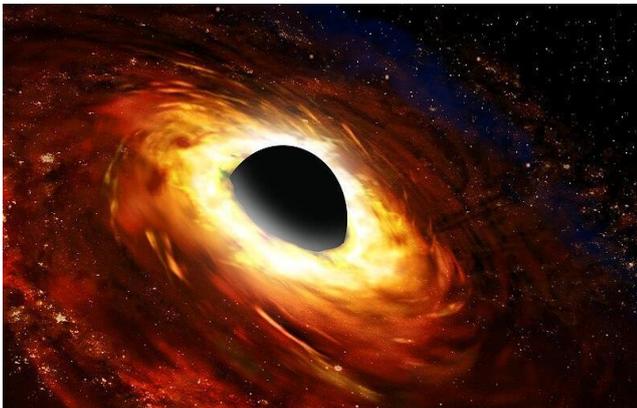

Fig.1. Artist's illustration of a BH surrounded by an AD (Source: https://tinyurl.com/8kw5whpp).

Figure 1 shows an artistic representation of a supermassive BH at the galaxy's center, surrounded by a hot, turbulent accretion disk (AD) of gas and dust drawn in by its powerful gravitational pull. The exact nature of AD viscosity remains elusive. To circumvent this, Shakura and Syunyaev [4] proposed a simplified model. Recognizing the insufficiency of molecular viscosity in inducing turbulence, Shakura and Syunyaev [4] introduced an effective viscosity parameter without delving into the complexities of turbulence theory. Additionally, an analytical solution for the AD model was presented by Pringle [5].

## II. IMPORTANCE OF SIMULATION IN ASTROPHYSICS

Accretion plays a crucial role in shaping the universe's formation and evolution, but its complex dynamics pose a significant challenge to understanding its mechanisms through theoretical frameworks alone, necessitating complementary approaches to unravel its intricacies. One critical aspect of accretion physics is developing and applying sophisticated numerical simulations (NS). Astrophysical phenomena, such as accretion, are governed by time-dependent flow dynamics equations, which must be solved numerically, as analytical solutions are often intractable, and this requires selecting the most efficient numerical method to ensure stable, computationally efficient and conservative solutions [6]. According to Giri [6], the equations governing accretion flows around BHs form a boundary value problem, which cannot be solved analytically, so numerical methods are used to study accretion, but to have confidence in the results, a solid understanding of analytical solutions is necessary. Numerical results should be verified against known analytical solutions whenever possible. NS are a powerful tool for exploring the dynamics of accretion flows around BHs and provide important details about BH's characteristics, evolution, and fundamental physical processes [7].

## III. THE SIMULATION TOOLKIT: A MULTIFACETED APPROACH

NS enhance our understanding of the complex behavior associated with accretion physics. Therefore, the purpose of this review paper is to provide an overview of some commonly used simulation methods that have successfully addressed the complex dynamic behavior of ADs.

*A. Hydrodynamic Simulations:*

To describe the fluid as a continuous medium with macroscopic properties such as pressure and density at each location, we can use hydrodynamic (HD) equations [8]. Equation of state (EoS) and appropriate boundary conditions, together with the conservation laws for mass, momentum, and energy, are to be used to describe the flow behavior collectively [9, 6]. Partial differential equations (PDEs) are solved numerically using a widely used technique called the finite difference (FD) method, which discretizes continuous flow variables (e.g., density, velocity) to certain approximative solutions [9, 6]. Similarly, because of its adaptability and flexibility, we can use the particle-based Smoothed Particle Hydrodynamics (SPH) method to simulate convection-dominated compressible flows [10]. SPH is a Lagrangian method in which the fluid is represented as packets of mass (termed particles) that solve HD equations in the framework of Lagrangian [9].

Giri [6] states that understanding the HD properties of BH accretion is essential because the spectral and temporal characteristics of the radiation emitted by matter flow and detected by satellites are directly influenced by the accretion flow's thermodynamic variables, such as temperature, energy density, and density.

*B. Magnetohydrodynamic Simulations:*

In astrophysical scenarios, fluids are typically highly ionized and conductive, frequently coexisting with magnetic fields. It is thus important to consider how fluid dynamics relates to embedded magnetic fields [11]. Magnetohydrodynamics (MHD) treats plasma as a continuous electrical conducting fluid and encompasses many processes in plasmas [12]. Many techniques are available for simulating MHD flows, including FD methods, finite element methods, finite-volume methods, and adaptive mesh refinement (AMR) techniques [13–16]. MHD simulations have yielded significant insight into the dynamical behavior and energetics of accretion flows in the presence of magnetic fields.

Recent advances in simulations of MHD turbulence in ADs have been substantial [8]. Simon et al. [17] and Davis et al. [18] used numerical MHD simulations to investigate turbulence in ADs. Many researchers have conducted NS to study the formation of magnetically collimated jets from BH ADs [19–21].

*C. Monte-Carlo Simulations:*

The Monte Carlo (MC) method is a numerical strategy that employs random variable simulations for solving mathematical problems [22]. Nicholas Metropolis and S. Ulam originally developed the MC technique at Los Alamos National Laboratory during the Manhattan Project to solve the Boltzmann equation for neutron distribution in fissile materials. The approach has since then been adopted in various scientific fields and widely used beyond its primary purpose [23].

When we encounter complex problems such as radiative transfer and Comptonization in accretion physics, we use the MC simulation technique. The radiation spectra emitted by the corona (electron cloud in an AD surrounding compact objects), like neutron stars and BH, can be simulated using the three-dimensional MC technique [23], since understanding the hot corona is crucial for studying BH ADs. Yao et al. [24] estimated the disk's inner radius and the corona size for the BH XTE J2012+381 by developing two models to represent different corona shapes using MC simulations.

The implementation of MC simulation is essential for understanding how a spectrum forms in compact bodies [25]. Ghosh [26] developed an MC radiative transfer (MCRT) code that the author applied to simulate radiative processes in advective accretion, demonstrating spectral variations with different parameters. According to Noebauer and Sim [22], the MCRT is a versatile approach that simulates the radiation processes and radiative transfer in astrophysics, linking physical properties to emitted radiation. The method requires managing MC noise, and various techniques are used to suppress it, with MCRT being applied to a range of astrophysical problems and showing promise for future developments.

## IV. SIMULATION CODES

Awesome astrophysical simulation codes: This is a list of astrophysical simulation codes grouped by functionality (https://github.com/pmocz/awesome-astrophysical-simulation-codes).

SPACE CoE: SPACE Centre of Excellence has a curated list of codes for astrophysical plasma simulations, including MHD codes for fluid dynamics (https://www.space-coe.eu/codes.php).

## V. IMPLICATIONS FOR ACCRETION PHYSICS

The HD simulations performed by Das and Sharma [27] led to a model for BH XRB transients, where changes in mass accretion rate can trigger transitions between hot and cold states, with different time scales corresponding to distinct physical processes. Das et al. [28] employed SPH to simulate the accretion of low-angular-momentum, viscous material onto BHs, expanding on earlier findings of shock formations and bipolar outflows in rotating transonic accretion flows. The results by Das et al. [28] show that increasing viscosity leads to time-dependent shocks, quasi-periodic variations in matter inflow and outflows, and modulated high-energy radiation, consistent with observed features in outbursting BH candidates.

According to the NS conducted by Massaglia et al. [29], the shape and structure of extragalactic radio sources are influenced by the specific properties of astrophysical jets, with lower power jets producing Fanaroff-Riley (FR) I-like plumes through turbulent dissipation and smooth deceleration, whereas 3D HD simulations are essential to capture this behavior. Again, 3D HD simulations by Gourgouliatos and Komissarov [30] have revealed that jets emitted by active galactic nuclei (AGN) experience a transformation into a turbulent state due to centrifugal instability, leading to reconfinement, which may be the underlying mechanism for the classification of jets into two distinct morphological categories according to the FR scheme. Sutherland and Bicknell [31] conducted 3D HD simulations to explore the interaction between a rapidly moving, low-density astrophysical jet and a dense, turbulent disk within an elliptical galaxy, applicable to extragalactic radio sources. According to Wagner et al. [32], relativistic astrophysical jets in radio galaxies drive outflows and star-formation through dense cloud collapse, with efficiency and importance influenced by interstellar medium properties and cloud distribution. Giri [6] conducted a study on the HD properties and stability of accretion flows using time-dependent NS. A custom FORTRAN code was developed to simulate 2D HD processes, leveraging the Total Variation Diminishing (TVD) method, a FD approach based on grid discretization, to model the behavior of viscous ADs surrounding BHs.

NS have been instrumental in understanding AGN jets since the 1970s [33]. The review by Martí [33] summarizes the current state of NS of AGN jets, covering their formation near the central BH and their impact on galactic scales, highlighting both achievements and limitations in interpreting jet phenomenology.

Viscous disks around BHs can form shock waves due to sub-Keplerian flow near the BH [34]. The simulations using SPH carried out by Lanzafame et al. [34] show that these shock waves influence the disk dynamics and radiation significantly, depending on the disk's properties and viscosity.

According to Komissarov and Porth [35], NS have led to significant advancements in understanding astrophysical jet formation, behavior, and interaction with surrounding plasma, with a promising outlook for future research.

Giri's article [36] reports on NS of transonic flow using HD models, exploring the effects of varying viscosity and cooling mechanisms. The simulations involve injecting sub-Keplerian matter at the outer boundary, which then accretes towards a BH. The results show a dual-structure solution, consisting of a thin, cool Keplerian disk near the equator and a transonic flow with shocks farther away from the equatorial plane, supporting the Two-Component

Advective Flow (TCAF) model proposed by Chakrabarti and Titarchuk.

Accretion onto BHs requires viscosity to remove angular momentum, with two dominant components (Keplerian and sub-Keplerian flows) whose dynamics are simulated numerically by Roy [7] to reproduce X-ray outbursts, hysteresis effects, and the formation/disappearance of Keplerian disks, shedding light on the complex accretion process.

It has been reported by Yuan [37] that numerical MHD simulations and observations agree that mass accretion rate decreases towards the BH. In an AD, the intense differential rotation rapidly stretches entangled magnetic field lines, transforming them into a predominantly toroidal configuration, particularly in weakly viscous, low angular momentum disks [38]. Simulations by Deb et al. [39] using a FD method show that toroidal flux tubes can significantly impact outflow properties, such as collimation and acceleration, and may even play a crucial role in shaping jets and outflows.

Many studies of NS have explored the development of standing shocks in low-angular-momentum, sub-Keplerian accretion flows around BHs, exploring the complex dynamics of accretion processes [40]. Singh et al. [40] employed the resistive MHD module of the PLUTO code to investigate 2D flow with low angular momentum around a BH. Simulations by Singh et al. [40] further show that flows with lower resistivity exhibit quasi-periodic oscillations (QPOs), while higher resistivity suppresses magnetic activity, resulting in a steady, symmetric flow.

Astrophysical objects often feature ADs and jets, but the underlying physical processes that govern their energy and angular momentum transfer remain incompletely understood. To shed light on this, Zanni et al. [41] conducted time-dependent simulations using the resistive MHD FLASH code with AMR, successfully modeling the launch of supersonic jets from magnetized ADs, exploring the effects of magnetic resistivity on the inflow-outflow system and seeking to establish steady-state solutions consistent with self-similar models. Fragile's review [42] provides an overview of the current state of knowledge regarding the processes involved in the formation of jets in BH ADs and magnetospheres through numerical MHD simulations, addressing key questions about jet properties and launching mechanisms and discussing the impact of initial simulation conditions and future research directions.

Accretion onto supermassive BHs (SMBHs) and subsequent jet formation is widely accepted as the power source behind AGNs, but the jet formation process, its relation to BH spin, and interaction with the surroundings remain debated topics [43]. Davis and Tchekhovskoy [43] have pointed out that NS, particularly general relativistic MHD and radiation MHD simulations, have advanced our understanding of jet production, disk stability, and the importance of magnetic fields, radiation viscosity, and spiral density waves in shaping AGN behavior.

The QPOs observed in BH candidates are thought to originate from oscillations in the Comptonizing regions of the accretion flow, which can be effectively modeled using the TCAF framework. NS conducted by Garain et al. [44] employing a TVD code combined with MC techniques validate this notion, demonstrating a correlation between QPO frequency, accretion rate, and spectral variations consistent with observational findings.

Bhattacharjee [45] presents a comprehensive review of theoretical models for neutron star accretion flows, synthesizing essential observations and modeling endeavors, and advocates for the suitability of the TCAF model in neutron star contexts, bolstered by supportive findings from MC and SPH simulations while also identifying avenues for future investigation.

A novel radiation HD simulation code was developed by Garain [46] to investigate the behavior of an AD surrounding a galactic BH, enabling the unprecedented simultaneous modeling of both rotating Keplerian and slower sub-Keplerian disk components. The study by Hilburn et al. [47] simulated Sagittarius A*'s (Sgr A*'s) accretion flow spectrum using coupled MC radiation and general relativistic MHD codes, comparing results to radio, infrared, and X-ray data.

Broderick [48] developed a theoretical model to explain the emission from magnetized plasmas accreting onto compact objects, considering polarization, gravitational, and Doppler effects. According to Broderick [48], relativity alters dispersion, polarized emission, and transfer effects in sheared flows, potentially generating circular polarization in XRBs, though not significant in AGN. The study by Broderick [48] develops a formalism for polarized radiative transfer through tangled magnetic fields, predicting frequency-dependent circular polarization due to magnetic helicity and potentially explaining observations in the Galactic center and XRBs.

A. Recent Findings through NS

The implementation of NS have led to numerous noteworthy findings in recent years. Among them, Garain et al. [49] used a 3D inviscid HD simulation technique to study the sub-Keplerian accretion flow dynamics onto a BH having no rotation. Their primary findings were the emergence of standing accretion shock instability (SASI) inside the AD. Nixon et al. [50] reported the numerical magnetic Reynolds numbers ($R_m$) in the dwarf nova disk using self-sustaining numerical modeling of dynamo dynamics and found lower values of $R_m$ by six orders of magnitude than the physically accepted value of $R_m \sim 10^{10}$. The accretion flow around SMBHs Sgr A* and M87* was specifically constructed using a radiatively inefficient, time-dependent accretion flow model in combination with 3D general relativistic magnetohydrodynamic (GRMHD) simulations by Chatterjee et al. [51]. The authors presented the first comparison of a sequence of increasingly complicated numerical models designed precisely to comprehend the accretion processes around the SMBHs Sgr A* and M87*. Donmez [52] used NS to provide a physical mechanism that explains several reported sources of QPO by applying a perturbation to the stable AD surrounding Einstein-gauss-bonnet (EGB) and Kerr BHs at varying angular velocity. They estimated that the EGB constant has a larger impact on QPOs than the BH spin parameter. Zhdankin et al. [53] investigated Rayleigh-Taylor instability using local kinetic particle-in-cell simulation with the Zeltron code in relativistic collisionless plasma and found efficient particle acceleration capable of explaining flares from inner accretion flows onto Sgr A* BH. Using 3D GRMHD simulation of an AD with high inclination around a fast-rotating BH, low angular momentum streamers that shower down on the inner sub disk driving accretion further were discovered by Kaaz et al. [54]. Jannaud et al. [55] carried out an investigation using NS of MHD jets from Keplerian ADs, and they reported the generation of standing recollimation shocks at great distances from the source for the first time. Having never been seen in earlier MHD simulations of jets, these recollimation shocks are essential to the MHD collimation process. These have been suggested as a natural consequence of the conditions for self-similar jet-launching [55].

Studies using NS on accretion are not limited to BHs or neutron stars. Recently, researchers have found many exciting aspects of planet formations from ADs using NS. In a radially organized protoplanetary disk, Jinno et al. [56] demonstrated that one can use N-body simulation to study the formation of planets via pebble accretion. Jinno et al. [56] used simulations to study the growth and inward migration of dust in the protoplanetary disk's dead zone region and investigate how planets form at the innermost boundary of the dead zone. Understanding external photoevaporation is essential to comprehend how most disks grow and, consequently, how most planets originate, according to Winter et al. [57]. Further, Winter et al. [57] reviewed the use of radiation HD simulation models to simulate external photoevaporation numerically. Additionally, it has been observed that paired planetesimal collisions are the predominant pathway of rocky planet accretion for the majority of parameter space corresponding to values of Shakura-Sunyaev turbulence parameters $(\alpha_v)$ reflected in observations of protoplanetary disks $\gtrsim 10^{-4}$ [58]. In that instance, Yap et al. [58] examined the genesis of planetesimal rings and the origin of the super-Earth by modeling the Stokes number in disks growing under MHD winds and turbulence.

## VI. SUMMARY

The field of accretion physics has come to rely heavily on NS as an essential tool for unraveling the intricate dynamics of the accretion process. However, NS have their limitations too. According to Malafarina [59], despite significant advancements in NS in the recent past, a fully satisfactory simulation of a supernova explosion resulting in BH formation remains an unresolved challenge. Van Box Som et al. [60] pointed out the shortcomings of one-dimensional simulations. They argued that multi-dimensional effects are essential to thoroughly understand the origins of QPOs and the non-linear dynamics of accretion columns in polars. Generally, two dimensions significantly challenge studying MHD turbulent flows [61]. A suitable resolution is yet another drawback. Proga [62] states that their model's primary drawback is its insufficient spatial resolution for simulating the magnetorotational instability inside the AD. Unlike local simulations, full global simulations aim to simulate a whole disk with varying radii, vertical stratification, and non-axisymmetric structure. Because of the vast scale of the global simulation, the physics that one may incorporate and the entire spatial resolution used must be limited [63].

High-performance computing is a must for NS. Astrophysicists heavily rely on supercomputers, but their problems demand even greater computational efficiency [9]. According to Klingenberg [9], future supercomputers will be massively parallel, requiring astrophysical numerical methods to adapt for optimal performance.

The article by Fragile [64] reviews the current state of NS of BH ADs, concentrating on large-scale simulations of the accretion process in the region immediately surrounding the BH within a radius of a few tens of gravitational radii. The review highlights exciting recent work in general relativistic MHD simulations, with some discussion of Newtonian radiation MHD and SPH, and looks to future directions in the field. While powerful computers and improved techniques have significantly boosted NS lately, it's likely that fundamental challenges in the field of accretion physics will persist for a long time [65].

ACKNOWLEDGMENT

The authors BS, LD, and AJB appreciate the reviewer's detailed comments and efforts to help improve the manuscript.